\documentclass[12pt]{article}

\usepackage{epsfig}
\usepackage{amssymb}
\title{Hydrodynamical Simulations of Clusters with {\it Galcons}}

\author{Yinon Arieli, Yoel Rephaeli \& Michael L. Norman}

\def\aa{{\it Astron. Astrophys.} \,}
\def\apj{{\it Ap. J.} \,}
\def\apjs{{\it Ap. J. Supp.} \,}

\def\apjl{{\it Ap. J. Lett.} \,}
\def\mn{{\it MNRAS} \,}
\def\aj{{\it Astron. J.} \,}
\def\ph{{\it astro-ph/}}
\def\na{{\it New Astron./}}

\def\aa{{\it Astron. Astrophys.} \,}
\def\apj{{\it ApJ.} \,}
\def\apjs{{\it Ap. J. Supp.} \,}

\def\apjl{{\it Ap. J. Lett.} \,}
\def\mn{{\it MNRAS} \,}
\def\aj{{\it Astron. J.} \,}

\def\rmp{{\it Rev. Mod. Phys.} \,}

\def\ph{{\it astro-ph/}}
\def\na{{\it New Astron. \,}}

\newcommand{\msun}{M_{\odot}}

\newcommand{\eg}{{\frenchspacing e.g.}}
\newcommand{\beq}{\begin{eqnarray}}
\newcommand{\eeq}{\end{eqnarray}}

\begin{document}
\title{Hydrodynamical Simulations of Galaxy Clusters with {\it Galcons}}

\newcommand{\D}{\displaystyle}
\pagenumbering{arabic}
\bibliographystyle{unsrt}
\maketitle


* accepted for publication in The Astrophysical Journal

\begin{abstract}

We present our recently developed {\em galcon} approach to hydrodynamical 
cosmological simulations of galaxy clusters - a subgrid model added to 
the {\em Enzo} adaptive mesh refinement code - which is capable of 
tracking galaxies within the cluster potential and following the 
feedback of their main baryonic processes. Galcons are physically 
extended galactic constructs within which baryonic processes are modeled 
analytically. By identifying galaxy halos and initializing galcons at 
high redshift ($z \sim 3$, well before most clusters virialize), we are 
able to follow the evolution of star formation, galactic winds, and 
ram-pressure stripping of interstellar media, along with their 
associated mass, metals and energy feedback into intracluster (IC) gas, 
which are deposited through a well-resolved spherical interface layer. 
Our approach is fully described and all results from 
initial simulations with the enhanced {\em Enzo-Galcon} code are 
presented. With a galactic star formation rate derived from the observed 
cosmic star formation density, our galcon simulation better reproduces 
the observed properties of IC gas, including the density, temperature, 
metallicity, and entropy profiles. By following the impact of a large 
number of galaxies on IC gas we explicitly demonstrate the advantages of 
this approach in producing a lower stellar fraction, a larger gas core 
radius, an isothermal temperature profile in the central cluster 
region, and a flatter metallicity gradient than in a standard simulation. 

\end{abstract}



\section{Introduction}
\label{sec:intro}

Hydrodynamical simulations of galaxy clusters, incorporating semi-analytic 
models for star formation and galactic feedback processes, show an appreciable 
level of inconsistency with observational results. This is particularly 
apparent in the simulated properties of intracluster (IC) gas - temperature, 
metallicity and entropy profiles - and the stellar mass density at high redshift 
(see the review by Borgani et al. 2008, and references therein). 
Statistical properties of clusters, such as X-ray luminosity-temperature, 
entropy-temperature, and mass-temperature relations, seem also to be discrepant 
when compared with high-precision optical and X-ray observations (e.g., Kay et 
al. 2007, Nagai et al. 2007, Tornatore et al. 2007, Kapferer et al. 2007; for 
a recent review, see Borgani et al. 2008). 
The mismatch between simulated and observed cluster properties (e.g., 
Evrard \& Henry 1991, 
Cavaliere, Menci \& Tozzi 1998, Tozzi \& Norman 2001) is largely due to 
insufficient 
accounting for essential physical processes, unrealistic 
simplifications of the evolution of star formation and feedback 
processes, and inadequate level of spatial resolution.

Some of the relevant physical processes that affect cluster properties and the 
dynamical and thermal state of IC gas are mergers of subclusters, galactic winds, 
ram-pressure stripping and gravitational drag. These have been partly implemented 
(e.g., Kapferer et al. 2005, Domainko et al. 2005, Bruggen \& Ruszkowski 2005, 
Sijacki \& Springel 2006, Kapferer et al. 2007) with some success in predicting 
IC gas properties. Different combinations of these processes and the various 
ways they are included in simulation codes generally result in quite different 
gas properties. An example is star formation (SF), whose self-consistent 
modeling (in cluster simulations) requires a prohibitively high level of spatial 
resolution which cannot be achieved with the current computing resources. Because 
of this limitation, most current simulations use a SF prescription that follows 
the formation of collisionless star `particles' in a running simulation 
(e.g., Cen \& Ostriker 1992; Nagai \& Kravtsov 2005), an approach which 
leads to an overestimation of the evolution of the SF rate (SFR) 
(Nagamine et al. 2004), and a higher than expected stellar to gas mass 
ratio. In addition, as we will show, in this particular implementation of SF the impact 
of the process remains spatially localized, resulting much lower mass 
(including metals) and energy ejection out of cluster 
galaxies, and consequently insufficient suppression of cooling and gas 
overdensity in cluster cores. This difficulty reflects the complexity of 
structure and SF processes, underlying the fact that a full 
implementation of these processes in hydrodynamical simulations is 
indeed a challenging task that nonetheless motivates attempts to 
develop a new approach in cluster simulations.

Considerations of galaxy clustering and star formation episodes at high redshift and the inclusion of 
heating to suppress gas cooling and condensation, lead us to identify Lyman 
Break Galaxies (LBGs) at $z \geq 3$ as early (`pre-heating') sources of IC gas. 
Implementation of longer episodes of SF in these galaxies induces stronger 
winds. As a result, the amount of energy and metal-rich gas ejected to IC 
space is higher, as required for consistency with observations. Gas dispersal 
is further enhanced by ram-pressure stripping. Incorporating these baryonic 
processes motivated us to develop a new approach in the description of the 
evolution of IC gas, one that is based on the powerful adaptive mesh refinement 
(AMR) cosmological hydrodynamical simulation code - {\em Enzo} (Bryan \& Norman 1997, O'Shea et al. 2004). We have modified and improved 
{\em Enzo} such that it is capable of following more realistically the 
hierarchical formation of structure through the 
inclusion of the most essential physical phenomena. This is accomplished by 
modeling the baryonic contents of galactic halos at high redshift by an extended `galaxy construct', which we refer to as {\it galcon}. The new 
{\em Enzo-Galcon} code does not require additional computational resources
compared to the original {\em Enzo} code. Because SF and feedback 
are modeled analytically, the level of resolution required to achieve 
improved results (compared to the standard simulation) is not extreme. 
This is the first time that most of the known processes are included in a hydrodynamical 
non-adiabatic simulation which is also capable of achieving high 
resolution ($\leq$ 10 kpc). Initial results from the first 
implementation of our galcon approach were briefly described by 
Arieli, Rephaeli \& Norman (2008, hereafter ARN).

In this paper a more complete description is given of our galcon 
approach, and an expanded analysis of the first simulations with the 
new code, including a wider range of IC gas properties than presented 
in ARN. In Section 2 we quantitatively describe the main baryonic processes 
included in our code. The galcon approach is introduced in Section 3, and 
results from the first {\em Enzo-Galcon} simulations are presented in Section 4, 
with a detailed comparison with the corresponding results from a (`standard') 
{\em Enzo} simulation using popular SF and feedback prescription. We end with 
a summary in Section 5.

\section{Gas Dispersal Processes}

The nearly cosmic metal abundances of IC gas attest to its largely 
galactic origin. Among the processes that were suggested to explain the 
ejection of metals form galaxies are winds (De Young 1978), ram-pressure 
stripping (Gunn \& Gott 1972), galaxy-galaxy interactions (Gnedin 1998, 
Kapferer et al. 2005), AGN outflows (Moll et al. 2006) and supernovae in 
IC space (Domainko et al. 2004). Observations seem to indicate that the 
two most important processes that transfer interstellar (IS) media to IC 
space, thereby building up the IC gas mass and its metal content, are 
galactic winds and ram-pressure stripping (e.g. Schindler 2008). In this 
section we describe the procedures for implementing these processes in 
our numerical code.

\subsection{Galactic wind}
\label{wind}

SN explosions generate shocks that drive metal-enriched galactic winds 
into IC space. The warm (shock-heated) IS media are further heated when 
winds from various galaxies merge and the velocities of the individual 
shocks are randomized (i.e., effectively thermalized). An observed 
morphological relation between the X-ray emitting gas and optical 
emission-line gas (Strickland et al 2000, 2002, Martin et al. 2002 \& 
Cecil et al. 2002) provides evidence for this process. The large width 
of these lines shows that accelerated IS material can reach velocities 
of several hundreds up to few thousands of km/s (Heckman et al. 2000). 
These and other X-ray observations (e.g., Sanders et al. 2004, De Grandi 
et al. 2004 \& Pratt et al. 2006) have observationally established the 
expected galactic origin of metals in IC gas, and that the mean 
metallicity is $Z \sim 0.3-0.4\, Z_{\odot}$, where $Z_{\odot}$ is 
solar metal abundance. 

The SN rate reflects the galactic formation rate of high-mass stars. 
Thus, the SF history is a basic feature of any model of energy and 
metallicity ejection out of cluster galaxies. Observations of galactic 
winds (\eg, Heckman 2003) provide direct evidence for the relation 
between mass and energy ejection rates and the SFR, $\dot{M}_\ast$,
\begin{eqnarray}
\label{eq:wind_ejecta}
\dot{E}_{w}=e_{w} \dot{M}_\ast c^2\; \nonumber \\
\dot{M}_{w}=\beta_{w} \dot{M}_\ast \;,
\end{eqnarray}
where the parameters $e_{w}$ and $\beta_{w}$ are energy and mass ejection 
efficiencies, respectively, and $c$ is the speed of light. These efficiency 
factors cannot be directly predicted from simple considerations, but they 
can be roughly estimated from observations (e.g., Pettini et al 2001, Heckman 
et al. 2001), from which typical values of $e_{w}=5\times 10^{-6}$ and 
$\beta_{w}=0.25$ are adopted (e.g., Cen \& Ostriker 1993, Leitherer et al. 
1992).

In the galcon approach, the SF history of each galaxy is deduced from 
observations and input to the simulation. This contrasts sharply with 
standard approaches for which the SF history is an output of the model. 
By fitting multiband photometric measurements with simulated 
galaxy spectra generated by a population synthesis model, the stellar mass 
density of galaxies can be deduced up to high redshift (Brinchmann \& Ellis 
2000, Cole et al. 2001, Cohen 2002, Dickinson et al. 2003, Fontana et al. 
2003, Glazebrook et al. 2004). Several groups have estimated the global 
stellar mass density in the redshift range $z=0-6$ by using this technique 
(Hopkins \& Beacom 2006, Nagamine et al. 2004, and references therein). 
The results strongly indicate that the SFR peaks at $z_{p} \simeq 3$, in 
accord with some semi-analytic models (e.g. Somerville et al. 2001, Menci 
et al. 2002), though in other works the SFR peaks at an earlier redshift, 
$z_{p} \geq 5$ (Hernquist \& Spirngel 2003, Nagamine et al. 2004). The 
global SFR density can be expressed in the form (Nagamine et al. 2006) 
\begin{equation}
\label{eq:SFR_fit}
\dot{\rho}_\ast = \dot{\rho}_0 \left[ t \,exp^{-t/\tau_d}/\tau_d + 
A\,t\,exp^{-t/\tau_s}/\tau_s \right]\;,
\end{equation}
where t is age of the universe, $\tau_d$ and $\tau_s$ are the 
characteristic SF times of disk and spheroid galaxies, respectively. 
The parameter A sets the scaling of the spheroid to disk stellar mass 
ratio, and $\dot{\rho}_0$ is determined by the overall normalization 
of the cosmic SFR. 
The mass density of galactic halos can be calculated using the Press 
\& Schechter (1974, hereafter PS) mass function 
\begin{equation}
\label{eq:PS}
\rho_g(z) = \int_{M_1}^{M_2} M n(M,z) dM\;,
\end{equation}
where $n(M,z)$ is the number density (per unit mass) of halos with a 
mass (in a differential interval around) M, and at redshift (interval 
around) z. The endpoints of the galactic mass range are taken to be 
$M_{1}=10^9 \,\msun$ and $M_{2}=10^{12}\,\msun$ as indicated by
observations (e.g., Nagamine et al. 2002).

We note that our use of the Press-Schechter mass function is for the 
sole purpose of gauging the star SF history in the two simulated 
clusters. While the Tinker et al. (2008) mass function is more 
consistent with results of current simulations, differences 
between these two functions are mostly important when describing 
statistical properties of the cluster population. For our main 
purpose here, namely a comparative study of the SF history in the 
CR and GR clusters, these differences are insignificant.

The SFR per unit galactic mass can be determined from the latter two 
expressions by simply taking their ratio, $s(z)=\dot{\rho}_{\ast}(z) / \rho_g(z)$. 
The total SFR in a galaxy with mass $M_g$ is 
\begin{equation}
\dot{M}_\ast(M,z) = M_g \,s(z)\;, 
\end{equation}
which we use to estimate energy and mass ejection by galactic winds. 
Note that we can readily consider also different forms for the function 
$s(z)$, reflecting SFR history which may be different in cluster galaxies 
than the cosmic average.

\subsection{Ram-pressure stripping}
\label{stripping}

During cluster collapse and ensuing episodes of galaxy and subcluster 
mergers, IS 
media are partly stripped by tidal interaction between galaxies. As 
IC gas density builds up, ram-pressure stripping becomes increasingly 
more effective, especially in the central, higher density region. We 
quantify this hydrodynamic process by determining for each galaxy (at 
any given time) the stripping radius, where local IC gas pressure is 
equal to the local galactic IS pressure. It is simply assumed that all 
IS gas outside this radius is stripped in a relatively short dynamical 
time.

Ram-pressure stripping (Gunn \& Gott 1972) was modeled using a simplified 
analytic expression (e.g., Domainko et al. 2006, Kapferer et al. 2007) 
which is valid for motion of a spiral galaxy in a direction perpendicular 
to its disk. The criterion for stripping was based on a comparison of the 
external pressure force exerted by IC gas with the gravitational attraction 
of the disk stars. This is clearly inappropriate in galaxies whose mass is 
dominated by DM, especially in the outer disk (and galactic halo). We thus 
generalize the stripping criterion at position R to include the gravitational 
force by the total galactic mass interior to R,
\begin{equation}
\label{eq:strip}
\rho_{ic} V_g^2 \pi R^2 = \frac{G M_{tot} M_s}{R^2} \; ,
\end{equation}
where $G$ is the gravitational constant, $\rho_{ic}$ is the local IC gas 
density, $V_g$ is the velocity of the galaxy in the cluster frame, and 
$M_s$ is the mass exterior to $R$. This relation provides an approximate 
estimate of the stripped mass, $M_s$, and the stripping radius, $R_s$, 
if the galactic gas density profile is known (see section \ref{model} 
for further details). 

Gas located outside the stripping radius is driven by ram pressure out 
of the galaxy. Stripping truncates the gaseous disk but is not expected 
to modify the profile of gas interior to $R_s$; neither does it appreciably 
affect the dynamics of the stellar and DM components of a galaxy (Kenny \& 
Koopmann 1999, Kenny, Van Gorkom \& Vollmer 2004).

\section{The {\em Enzo}-Galcon code}

Various SF prescriptions have been used to form new collisionless star 
`particles' in cosmological simulations (e.g., Cen \& Ostriker 1992; 
Katz, Weinberg, \& Hernquist 1996, Springel \& Hernquist 2003; Nagai \& 
Kravtsov 2005). The basic approach 
in these recipes is that star groups are formed only in regions with 
density higher than some constant threshold density. However, current 
cosmological simulations are limited to a level of spatial resolution 
of not less than a few kpc, whereas star groups are formed on much 
smaller length scales (10-100 pc). Furthermore, simulations are also 
limited by mass resolution, thus only large star clusters can be formed, 
a result which is inconsistent with observations of much smaller star 
groups. To be more realistic, simulations need to attain higher spatial 
and mass resolutions, which obviously necessitate prohibitively high 
computing resources (for this task). Other related physical processes, 
such as galactic winds and ram-pressure stripping, are sufficiently 
complex and cannot be realistically modeled using the current heuristic 
SF and feedback mechanisms.

Current cluster simulations also fail to preserve galaxy-size halos due 
to substantial overmerging during cluster formation (Klypin et al. 1999). In practice, even when 
using higher spatial and mass resolutions this problem is only somewhat less 
severe, and simulated clusters still have only a small number of identified 
galaxies. In principle, AMR codes are capable of preserving the less massive 
halos from overmerging if they solve the dynamics of DM and gas in the 
relevant volume with an adequate level of refinement. However, halos are not 
pre-defined objects during the run, and their locations and sizes can only be 
deduced efficiently off-line. Thus, the volume that needs to be refined at 
each time step is unknown, making this approach impractical to perform in 
current AMR codes. These drawbacks are probably the main reasons for the 
discrepancy between results of cosmological simulations and observations 
(e.g., Borgani et al. 2008).

The inadequacy of spatial and mass resolution in current 
cosmological simulation codes poses a challenge that motivated us to 
seek a new direction in attaining an improved description of the 
evolution of clusters. More specifically, the fact that our main interest 
is the evolution of IC gas - rather than an improved numerical description 
of galactic structure - led us to develop a semi-analytic model for the 
galaxy as a whole. We replace the baryonic contents of the under-resolved 
galaxies with a physically extended galaxy subgrid model - a galcon. As 
is demonstrated below, the introduction of galcons allows us to substantially 
simplify analytic modeling of SF, wind, and ram-pressure stripping processes. 
Galcons are initialized at the center of DM halos, and the implementation 
of these processes is affected by simple analytic models, allowing full 
control of the evolution of the stellar and gaseous contents of galaxies. 
Due to their extended morphology, mass and energy ejecta from galcons are 
deposited at large galactic radii, directly mixing there with IC gas. 
This prescription is very different from previous models in which ejecta 
were deposited at the galactic center, with little likelihood of ever 
leaving the galaxy. Moreover, since galcons are fully identified 
systems, their positions and dimensions are known at each time step. 
This allows us to use a second refinement criterion, in 
addition to the regular density refinement criterion, which refines 
cells that contain DM or gas densities higher than a given value. 
The other criterion is geometrical: At each timestep cells that are 
placed inside the outer radius of each galcon, and those that 
are located in a layer surrounding each galcon, are refined to the 
highest attainable level of resolution. This ensures that all halos 
are adequately resolved, improving the description of both the 
dynamics and hydrodynamics in these regions, thereby preventing the 
overmerging of halos at later times. This seems to be the first time 
that a geometrical refinement criterion based on simulation object 
is used in a hydrodynamical simulation.

Our cluster simulations are based on the powerful {\em Enzo} adaptive 
mesh refinement (AMR) code (Bryan \& Norman 1997; Bryan 1999; Norman \& 
Bryan 1999; Bryan, Abel \& Norman 2001, O'Shea et al. 2004). {\em Enzo} 
is a grid-based hybrid code (hydro + N-body) which uses the AMR 
algorithm of Berger \& Collela (1989) to improve spatial resolution 
in regions of large gradients, such as in gravitationally 
collapsing objects. The method is attractive for cosmological 
applications because (1) it is spatially-and 
time-adaptive, (2) it uses accurate and well-tested grid-based methods 
for solving the equations of hydrodynamics, and (3) it can be well 
optimized and parallelized. The central idea behind AMR is to solve the 
evolution equations on a grid, adding finer meshes in regions that 
require enhanced resolution. Mesh refinement can be continued to an 
arbitrary level, based on criteria involving any combination of 
overdensity (DM and/or baryon), Jeans length, cooling time, etc., 
enabling us to tailor the adaptivity to the problem of interest. The 
code can follow the evolution of the following physics models: 
Collisionless DM using the particle-mesh N-body technique; gravity, 
using FFTs on the root grid and multigrid relaxation on the subgrids; 
cosmic expansion; gas dynamics, using the piecewise parabolic method 
(PPM); multispecies nonequilibrium ionization and H2 chemistry, using 
backward Euler time differencing, and radiative heating and cooling, 
using subcycled forward Euler time differencing.

Customizing {\em Enzo} with galcons enables us to follow the hierarchical 
formation of a cluster, and description of its DM and IC gas properties, 
more realistically by including the most important physical phenomena, 
yet attaining a high level of spatial resolution even in the central region 
of the cluster. This is achieved by performing a high resolution cosmological 
simulation, starting at some initial redshift (usually $z \simeq 60$) and 
proceeding up to a redshift $z_r$ in the range $z=3-5$. At $z_r$ galactic 
DM halos, basic elements of the protocluster, are located by one of the 
halo finding techniques (Eisenstein \& Hut 1998), and the baryon density 
profiles are examined. Then, the baryonic mass inside these halos is 
replaced by a galcon with the same density profile and total mass (so 
that an unphysical instantaneous change in the simulated density 
distribution does not occur). After the replacement, the simulation is 
continued to redshift zero, with galcons traveling in a sea of DM 
particles and gas, enriching IC gas with energy, mass, and metals. Note 
that Metzler \& Evrard (1994) had already attempted to incorporate 
galactic winds in a simplified description of galaxies, but in their 
SPH simulations galaxies were treated as point particles, and the wind  
prescription was overly simplified.

\subsection{Models}
\label{model}

A key aspect of our approach is based on SF and feedback in galcons. 
Strong observational evidence that the SFR peaked at the redshift range 
$z \sim 3-5$, combined with the knowledge that high-redshift galaxies 
were then already well developed, lead us to introduce galcons at this 
epoch of protocluster evolution. A hydrodynamical non-adiabatic {\em Enzo} 
simulation is evolved with no SF until a redshift at the above range; 
at this replacement redshift, the run is stopped and galactic halos with 
total mass in the range $10^9-10^{12} M_\odot$ are identified. This is the 
mass range of high redshift galaxies, mainly LBG, as indicated by 
observations (e.g., Nagamine et al. 2002). 

The baryonic density profile of each halo is truncated at $r_{200}$, the 
radius where the density is 200 times the background level, and the 
baryonic content is then analyzed. First, the outer radius ($R_M$) 
of the region containing 90\% of baryonic mass is determined; this 
will be the radius of the region where feedback energy and mass are 
deposited. This choice of radius ensures that most of the galaxy baryonic 
gas is included in the galcon, while avoiding discontinuity in the mass 
profile when inserting the galcon into the galactic center. The gas mass 
density is then fit by a $\beta$-profile, $\rho_b=\rho_{0}/[1+(r/r_c)^2]^
{-3\beta/2}]$, where $r_c$ is a core radius, and $\rho_0$ is the central 
density. This profile describes well the baryonic contents of high 
redshift galaxies. 

Both stellar and gaseous components are included in galcons, and since 
their spatial distributions are expected to be initially very similar, 
it is reasonable to take for both the same $\beta$-profile parameters, 
but with different central densities. Adding up the stellar and gaseous 
contents of each halo yields the total baryonic mass, $M_b$, and the 
initial central galcon baryon density 
\begin{equation}
\rho_0=\frac{M_b}{4 \pi r_0^3 \int_0^{R_{M}/r_0} \frac{x^2 dx}{\left(1+
x^2 \right)^{3\beta}/2}} \;,
\end{equation}
where $x=r/r_c$. Initialized with the velocity of the halo, the galcon 
is inserted into the center of the halo and initialized with an extended 
spherical density distribution using the fitting parameters that were 
found earlier. 

To determine the initial gaseous-to-stellar mass ratio, $f_{g-s}$, 
at $z_{r}$ the gas and stellar mass densities have to be calculated. 
The total baryonic mass density can be evaluated by multiplying the 
mass derived from the Press \& Schechter (1974) mass function by the 
universal baryonic density $\Omega_b$
\begin{equation}
\rho_b(z) = \Omega_b \int_{M_1}^{M_2} M n(M,z) dM\;,
\end{equation}
where $n(M,z)$ is the number density of halos with a given mass and 
redshift, and the integration limits are the same as in Eq. \ref{eq:PS}.
The star mass density can be evaluated by integrating the cosmic SFR 
density from Eq. \ref{eq:SFR_fit}
\begin{equation}
\rho_s(z=z_{r})=\int_{t_s}^{t_{r}} \dot{\rho}_\ast dt \;,
\end{equation}
where $t_s$ and $t_{r}$ are the times when SF began and the replacement 
time, respectively. Since the cosmic SFR density dramatically decreases at 
$z \geq 6$, we simplify somewhat by taking $z_s = 6$.

The gas-to-star mass ratio in galaxies at $z_r$ can now be determined
\begin{equation}
f_{g-s}(z_{r})=\frac{\rho_b(z_{r})-\rho_s}{\rho_s}\;.
\end{equation}
Taking $z_s$=6, and assuming that the peak in the cosmic SFR density is 
at $z=3$, we compute $f_{g-s}(z_{r} \sim 3)$. This result is used to 
set the initial central density of the galcon stellar and gaseous 
components such that its extended density distribution has the form
\begin{equation}
\rho_{G}=\frac{\rho_{s0}+\rho_{g0}}{\left[ 1+\left(r/r_0 \right)^2 
\right]^{3\beta/2}} \;,
\end{equation}
where $\rho_{s0}=\rho_{g0}/f_{g-s}(z_{r})$ and $\rho_{g0}$ are the 
stellar and gas central densities. 

Having initiated galcons as described above, we can now follow the mass 
and energy ejection processes that enrich IC gas - galactic 
winds and ram-pressure stripping. While both processes reduce the mass 
of IS gas, only winds reduce the total mass of (already formed) stars.
Winds affect only the stellar component of galcons by reducing the stellar 
mass. Ram pressure stripping continuously reduces the galcon outer gas 
radius. Because galactic winds are SN driven, their elemental abundances 
are higher than in IS gas, by a factor of $\sim 3$. The enrichment by 
each of these processes is separately followed.

Galcons are treated in the code as collisionless particles, but with an 
extended density distribution as described earlier. In the current phase 
of the work we do not include any enhancement in the SFR due to mergers 
of galcons. However, since the code includes a full treatment of the 
dynamics of DM, galcons, and density fields, mergers do occur in the 
simulation. An explicit description of mergers and their impact will be 
studied in the second phase of our work. 

At each time step ($\Delta t$) the code calculates the current SFR of 
each galcon according to eq. \ref{eq:SFR_fit}, based on which the energy 
and mass ejected by the wind are determined by 
\begin{eqnarray}
\Delta E_{w}&=&e_{w} \dot{M}_\ast c^2 \Delta t \nonumber\\
\Delta M_{w}&=&\beta_{w} \dot{M}_\ast \Delta t\;.
\end{eqnarray}
The ejected stellar mass contains a blend of heavy metals which 
incrementally increases the metallicity of IC gas by an amount which is 
proportional to galcon mass ejecta. The transfer of mass and energy to IC 
gas is implemented by isotropically distributing the ejecta over a thin 
layer (which typically has the size of the maximal resolution length) 
surrounding the outer galcon radius. Note though that the dependence on 
the thickness of this layer is relatively weak since we achieve a 
resolution of 9 kpc, obviously much smaller than a typical IC 
gas core radius of $200-300$ kpc. In our initial work outflows are 
taken to be isotropic; this assumption can be relaxed in follow-up work 
to include the possibility of bipolar or some other non-isotropic flow. 

Accurate determination of the metallicity of the ejecta requires a 
detailed description of the initial mass function (IMF), the SF 
process, and metal yields of different SN types. Since the primary 
objective of this work is to investigate the overall impact of the 
galactic feedback processes on IC gas, irrespective of the detailed 
stellar processes involved in ejecting metals into IS gas, we simplify 
our model by considering the overall SFR from the entire disk (as 
described in section \ref{wind}). 

SN ejecta from the galactic disk quickly interact through shocks
and mix with the surrounding IS gas, so that the wind contains a blend
of heavy metals from stars and IS gas. However, the fraction of gas
in the ejecta cannot be accurately determined. We assume that most of
the wind ejecta come from the stellar component whose metallicity is
approximately solar; accordingly, we subtract the energy and mass that
is carried by the wind from the galcon stellar content and increase
the metallicity of IC gas by an amount which is proportional to the 
galcon mass ejecta. Also, since the level of SN activity is roughly 
linearly proportional to the local star density, the wind does not 
modify the spatial stellar profile. Thus, we only adjust the central 
density of the stellar component to reflect the loss of stellar material, 
\begin{equation}
\rho'_{s0}=\rho_{s0} \frac{M'_{s}}{M_{s}} \;,
\end{equation}
where $M'_{s}=M_s-(1-\beta_w) \dot{M}_\ast \Delta t$, and $M_s$ 
are the current galcon stellar mass and its value in the previous time 
step, respectively.

As discussed in section \ref{stripping}, stripping is modeled by ejecting 
the gaseous mass located outside of the stripping radius into IC gas. Eq. 
\ref{eq:strip} is used to compute the mass to be stripped at the current 
time step 
\begin{equation}
\label{eq:strip_mass}
M_s=\frac{\rho_{ic} V_g^2 \pi r_i^4}{G M_{tot}(r_i)} \;,
\end{equation}
where $r_i$ is the outer radius of the gaseous sphere at the current time 
step. Knowing the radial ($\beta$) profile of the gas, this mass 
is used to find the stripping radius $r_i$. Gas located outside this 
radius is then removed from the galcon and deposited isotropically into 
IC space. As noted already, stripping only truncates the gas outer radius 
without modifying its mass profile; thus, the outer radius of the galcon 
gas component is now truncated at the stripping radius, but the scale 
radius and central density are unchanged.

\section{First Galcon simulations}
\label{results}

We have performed two sets of high resolution simulations with AMR and 
radiative cooling (Sutherland \& Dopita 1993). The first - referred to 
as the galcon run (GR) - included the above physical processes and the 
galcon algorithm, whereas the second, an {\em Enzo} run with no galcons - 
the comparison run (CR) - included only the SF and feedback recipe of Cen 
\& Ostriker (1992). We carry out a detailed comparison of results from these 
two runs in order to quantify the improvements in the description of the 
evolution of IC gas in our new galcon code. The root grid in both runs 
includes $128^3$ cells which cover a comoving volume of $54^3 \, Mpc^3$ 
with two nested inner grids. The highest refined grid covers a comoving 
volume of $27^3 \, Mpc^3$ divided into $128^3$ cells; this can be further refined by 
up to 5 levels, with a maximum $\sim$9 kpc resolution.

The simulations were initialized at $z=60$ assuming a $\Lambda$CDM model 
with $\Omega_m=0.27$, $\Omega_\Lambda=0.73$, $\sigma_8=0.9$, and 
$h=0.71$ ($H_0$ in units of 100 $km\; s^{-1}\; Mpc^{-1}$). The CR 
run was evolved continuously to $z=0$. The GR run was stopped at $z=3$, 
and a halo-finding algorithm (Eisenstein \& Hut 1998) was used to locate 
89 galactic halos with mass in the range $10^9-10^{12}\; M_\odot$ 
within a volume which eventually collapsed to form the cluster. The 
baryonic contents of these halos were analyzed and replaced by galcons, 
as described in section \ref{model}. The GR run was then resumed and 
evolved to $z=0$ with the additional physical processes as described in 
the previous section. Results from these two runs are presented in the 
next two subsections.

\subsection{Basic cluster properties}
\label{cluster_properties}

The two simulated (GR and CR) clusters are rich, $\sim 5.4 \times 
10^{14} \; M_\odot$, with very similar global properties, which are 
summarized in Table \ref{table:cluster}. Clearly, the near equality in 
the global properties of the clusters stems from the identically followed 
DM dynamics which govern cluster formation and evolution. Significant 
differences are expected and seen in the the baryonic mass components, 
particularly in the central regions. 

First, a significant difference is seen in the number of cluster galaxies; 
within the virial radius of the GR cluster 89 galaxies were identified 
(Fig. 1), as compared with only 6 galaxies in the CR cluster. The 
drastically lower number of identified galaxies in the CR cluster is a 
result of inadequate force resolution and consequently unphysical merging 
of galaxy DM halos (the ``overmerging problem"; Moore et al. 1996, Klypin 
et al. 1999.). It is argued that a {\em proper} force resolution of 
$\leq 2 h^{-1}$ kpc, and mass resolution $\leq 10^9 h^{-1} 
M_{\odot}$ are required for galaxy mass halos to survive in the dense 
cluster core (Klypin et al. 1999). In our {\em Enzo} simulations this 
mass resolution requirement is met, but the 9 kpc {\em comoving} 
force resolution is apparently too coarse for preventing 
overmerging. On the other hand, by replacing the baryon contents of 
galaxies with galcons at z=3, when the proper force resolution is 
four times better - and the mass distribution determined to a higher 
degree of detail - we effectively force retention of `memory' of this 
state in latter stages of the cluster evolution in a way that is 
essentially resolution-independent. During the N-body dynamics phase of 
the calculation, the extended mass distribution of galcons is deposited to
the mesh where it helps anchor the DM halo, despite less than optimal 
force resolution. Even so, some galcons did merge, but their number 
remained much higher than in the CR cluster whose halos merged 
successively to form higher mass systems. The fact that the galcon 
approach requires substantially lower resolution for overcoming the 
overmerging problem constitutes a significant advantage over other 
simulation codes, an advantage that has important consequences for IC 
gas properties, as demonstrated below.

\begin{table}
\label{table:cluster}
\begin{center}
\begin{tabular}{|c|c|c|} \hline\hline
             & GR cluster & CR cluster \\\cline{1-3}
Total Mass   & 5.44 & 5.43 \\\cline{1-3}
DM mass      & 4.89 & 4.88 \\\cline{1-3}
Gas mass     & 0.504 & 0.434 \\\cline{1-3}
Stellar mass & 0.045 & 0.106 \\\cline{1-3}
$R_{vir}$ & 1.71 & 1.71 \\\cline{1-3}
\end{tabular}
\end{center}
\caption{Main properties of the clusters identified in the GR and CR 
simulations. Masses are in units of $10^{14} M_\odot$, and radii in 
Mpc.}
\end{table}

\label{fig:GPS_locations}
\begin{figure}[h]
\centering
\epsfig{file=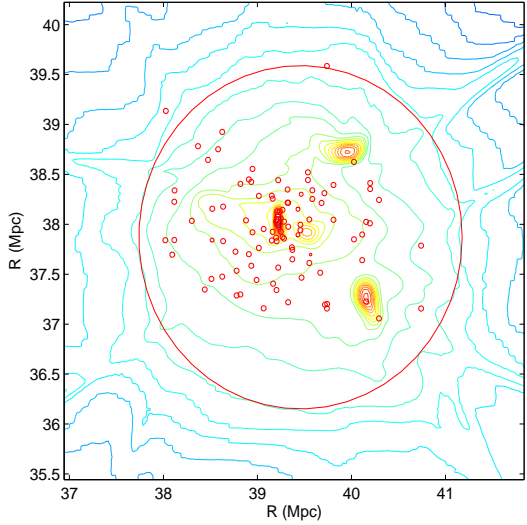,height=7cm,width=7cm,clip=}
\caption{Locations of the galcons in the GR cluster (small circles) 
plotted on the density contour map. The cluster virial radius is 
indicated by the large circle.}
\end{figure}

\subsection{DM and galcon velocity distributions}
\label{sec:beta}

While the main focus of our first galcon simulations have been IC gas 
properties, it is very much of interest to compare the galcon velocity 
distribution with that of DM. The relatively large number of identified 
galcons makes it feasible to determine the velocity anisotropy profile also 
for the cluster galaxies in our GR simulation. Since galcons are essentially 
collisionless, their kinematics would be expected to resemble that 
of DM. Any differences in the phase space distributions of galaxies and 
DM, especially in the central region where galaxy density is high and 
- under the action of gravitational drag (e.g., Rephaeli \& Salpeter 
1980) - the galaxy distribution may show some indication of a weak 
degree of collisional evolution, which could be manifested in a 
faster approach to isotropy. For this and other reasons it is of interest 
to compare the velocity anisotropy of these two matter components 
\begin{equation}
\beta_v \equiv 1-\frac{\sigma_t^2}{\sigma_r^2},
\end{equation}
where $\sigma_t^2$ and $\sigma_r^2$ are the 1-dimensional tangential 
and radial velocity dispersions (in a spatially spherically symmetric 
system). The late assemblage of clusters, and their relatively long 
crossing times, would suggest that (even DM) velocities are not likely 
to isotropize in the outer regions of clusters where matter components 
(especially recently accreted DM and galaxies) are not well-mixed. 

\begin{figure}[h]
\label{fig:beta_dm_galcons}
\centering
\epsfig{file=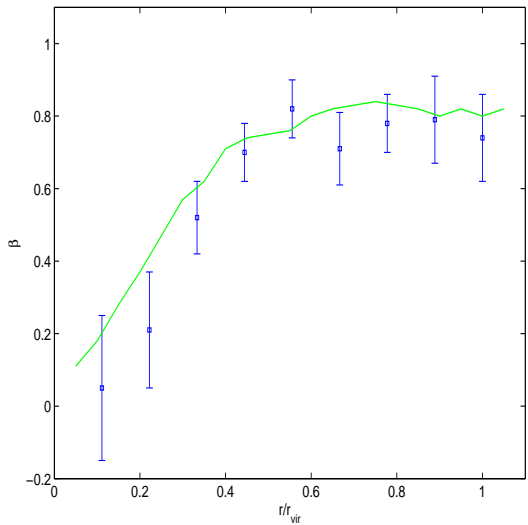,height=7cm,width=7cm,clip=}
\caption{The velocity anisotropy of DM (solid green line)
and galaxies (blue squares with error bars) in the GR cluster.}
\end{figure}

The spatial distributions of DM and galaxy velocities in clusters are of 
great interest for characterization of cluster morphology and dynamical 
evolution. For example, the comparison of velocity anisotropy profiles of 
these two (mostly) collisionless components can reveal details of their true 
phase space character. The ability to investigate this and other issues has 
improved significantly with the more extensive observational databases that 
are now available - large samples of both positions and velocities of member 
galaxies, DM and total mass profiles deduced from joint gravitational lensing 
and X-ray maps. In a recent work, Host et al. (2009) used X-ray observables of 
IC gas for inferring the radial profile of the velocity anisotropy of the cluster 
DM based on assuming a universal relation between DM velocity dispersion and 
gas temperature, calibrated using numerical simulations. Applying this approach 
to both low and intermediate redshift samples, they found that the DM velocity 
anisotropy is significantly different than zero, indicating that DM is effectively 
collisionless. Lemze et al. (2009) used a model-independent method to solve the 
Jeans equation, simultaneously incorporating the observed velocity dispersion 
profile and galaxy number counts, to study the velocity anisotropy of the 
cluster Abell 1689. In accord with previous results, velocities were found 
to be mostly radial in the outer region of the cluster and mostly tangential in 
the central region.

Numerical N-body simulations of collisionless DM also show that velocities are 
anisotropic in the outer region, with $\beta_{DM}$ greater than zero in the 
central region, and increasing to $\sim 0.8-1$ near the virial radius (e.g., 
Colin et al. 2000, Hansen \& Stadel 2006, Host et al. 2009). Similar 
behavior is seen in figure 2, where $\beta_{DM}$ is plotted as 
function of radial distance from the 
center of our simulated GR cluster. The corresponding values of 
$\beta_v$ for the galcons, shown by the squares (with their associated 
$1-\sigma$ uncertainty intervals), follow the DM curve above $0.5 
r_{vir}$. There is some indication that galcon velocities are somewhat 
more isotropic in the central region, a trend that is in accord with 
theoretical expectation. It will be interesting to see if this behavior 
will also be found in future simulations with a larger number of galcons. 

Also indicative of the degree of relaxation is the smoothness of the anisotropy 
profile. The presence of significant sub-clumps and ongoing merger activity would 
likely result in a non-smooth $\beta_v$ profile; as can be seen in figure 
2, the GR cluster has a relatively smooth profile 
indicating that the cluster is well relaxed. These preliminary results 
demonstrate the enhanced capability of probing the structure and 
evolution of clusters with our {\em Enzo-Galcon} code.

\subsection{Star formation and heating}
\label{sec:star-formation}

SF history is a central driver of gas feedback processes. It is therefore 
very important to verify that our GR simulation yields an acceptable stellar 
mass fraction, and produces the required gas heating to overcome overcooling. 
As stated before, we adopt the functional redshift behavior of the 
observationally deduced cosmic SFR (see Eq. \ref{eq:SFR_fit}) to 
calculate the SFR of each galcon at a given z. 
However, each galcon has a different initial mass, which varies during 
the evolution of the galaxy it represents. Therefore, SFR in galcons is 
calculated based on the total mass of its parent halo at each redshift, 
and is only indirectly determined by the global cosmic SFR. 
Figure 3 shows a comparison 
between the resultant cosmic SFR in the GR and CR clusters with the 
observational data. 

\begin{figure}[h]
\label{fig:SFR_result}
\centering
\epsfig{file=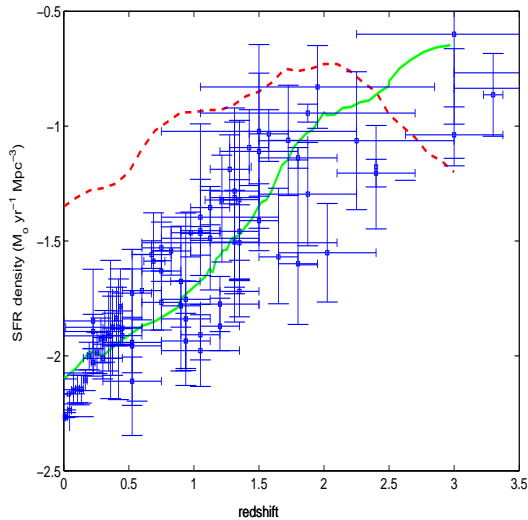,height=7cm,width=7cm,clip=}
\caption{A comparison of the resultant cosmic SFR density in the GR 
cluster, solid green line, and the CR cluster, dashed red line, with 
observational data (Hopkins \& Beacom 2006 and references therein).}
\end{figure}

As is quite obvious from Figure 3, the cosmic SFR density of the GR cluster 
is consistent with the observational data at all redshifts (in the interval 
considered here). The resultant SFR density in the CR cluster behaves 
differently and does not fit well the observations. The SFR density is 
lower than the observed values at higher redshift and higher at lower 
redshifts. Moreover, the peak of the SFR density is around $z \sim 2$, 
significantly later than the observationally deduced peak 
redshift of $z=3$. In the CR simulation the prescription 
of Cen \& Ostriker (1992) is implemented to form new star particles. 
Generally, groups of stars are formed in regions where the density 
is higher than a threshold value, or when cooling times are shorter than a 
characteristic timescale. As mentioned earlier, accurately resolving the full 
cluster volume requires prohibitively large computing resources. Therefore, 
it is likely that gas clumps are not treated realistically, leading to 
very imprecise estimation of the stellar mass. This can be assessed by 
calculating the volume over which feedback occurs at each redshift. As 
can be seen from Figure 4, the feedback process in the CR 
cluster is limited to a smaller region than in the GR cluster 
at all redshifts. Consequently, the CR cluster contains a higher fraction 
of cool gas, and therefore also increased SFR and stellar mass. The 
enhanced SFR leads to a high star-to-gas mass ratio of $\sim$22\%, about 
twice higher than the observationally deduced value (e.g., Balogh et al. 
2001; Wu \& Xue 2002).

\begin{figure}[h]
\label{fig:local}
\centering
\epsfig{file=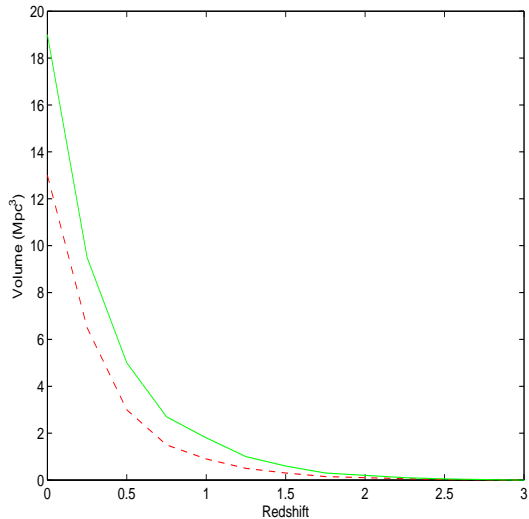,height=7cm,width=7cm,clip=}
\caption{Feedback volume at different redshifts. A comparison of the volumes 
over which feedback occurs in the GR cluster, shown by the solid green line, 
and the CR cluster, shown by the dashed red line.}
\end{figure}

On the other hand, energy feedback due to wind in the GR cluster succeeds 
in supplying a sufficient amount of thermal energy to heat the cold gas, 
thereby preventing the formation of too many stars. As a result, mass of 
the cold component is only $\sim$9.5\%, as compared to the hot IC gas mass. 
In contrast with the localized nature of SF and feedback in the CR cluster, 
in the galcon approach SF is much less local, and energy is deposited directly 
to a region immediately outside the galaxy. Thus, energy spreads out more 
easily, heating IC gas more efficiently. 
In addition, the strength of the feedback in the galcon approach is directly 
determined by the mass of the galaxy, since the feedback is proportional to the 
SFR. Thus, as required, feedback is enhanced in the cluster central region. 
These two features underline the significant differences between the two 
respective SF models. The capability of the galcon approach to correctly 
reproduce both the proper cosmic SFR history and the amount of stars in the 
cluster highlights its advantages, and demonstrates the importance of a more 
realistic description of the the relevant physical processes.

\subsection{Gas density and temperature}

Clearly, IC gas properties are strongly affected by the nature of physical 
processes that occurred during cluster collapse and evolution. In Figure 
5 IC gas density profiles in the two clusters are 
compared. The density profiles are similar at large radii, including a steep 
hump at $r \sim$600kpc, indicating the location of a very massive clump. The 
profiles flatten toward the center, but the GR cluster has a substantially 
larger core of $\sim$180 kpc, as compared to a relatively small core of $50$ 
kpc in the CR cluster. Gas core radii of massive isothermal clusters, similar 
to our simulated clusters, are typically in $200 - 300$ kpc (e.g., Sun M. 
et al. 2009).  While the core of the GR cluster is at the low end of this 
range, the core of the CR cluster unrealistically small. The shape of 
the density profile in the central region is mainly determined 
by the fraction of IC gas that cooled down and converted into stars. The 
absence of sufficient feedback in the CR cluster results in excessive 
mass of cool gas in the inner core. Strong cooling results in a small 
core as well as an unrealistically high number of stars, whereas in the 
GR cluster feedback is stronger and more efficiently spread, resulting 
in suppression of overcooling, and in a larger core.

\label{fig:density}
\begin{figure}[h]
\centering
\epsfig{file=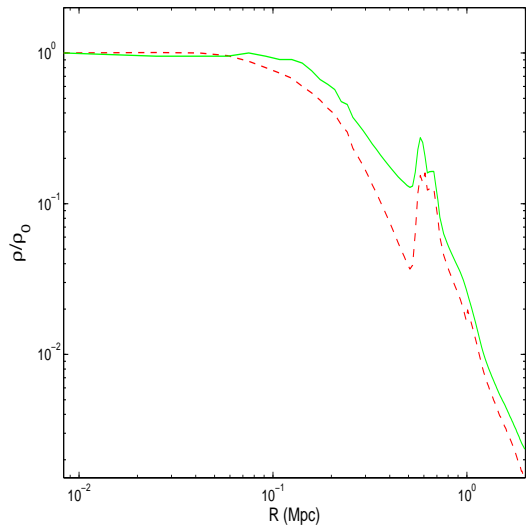,height=7cm,width=7cm,clip=}
\caption{Normalized density profiles of IC gas in the GR cluster, 
solid green line, and the CR cluster, dashed red line.}
\end{figure}

Other baryonic properties differ substantially in the two clusters; 
perhaps the most prominent difference is shown in the temperature 
distribution, which is plotted in figure 6. Neither 
the clusters has a cool core. Non cooling-flow clusters have an 
isothermal flat core and the temperature profile declines rapidly 
outside the core (e.g., Snowden et al. 2008, Pratt et al. 2007). 
As can be seen in figure 6, both clusters do show a rapid decline 
in the temperature profile at radii larger than $\sim$200 kpc. 
However, only the GR cluster has a flat isothermal core, while in 
the CR cluster the temperature continues a moderate rise toward the center. 
The difference between the two profiles is more pronounced when compared to 
observational results. In figure 6 we show the mean temperature profile 
with its $1\sigma$ uncertainty region as deduced from XMM-Newton 
observations of 15 nearby clusters (Pratt et al. 2007). The mean profile 
of this sample shows an isothermal behavior in the central region, 
although it slightly decreases due to 4 cool-core clusters that are 
also included in the sample. 

In the CR cluster a local SF prescription is used to create star groups in 
cells with high density and low temperature. The feedback from these star 
groups enriches and heats their immediate surroundings, spreading only 
over a small central region. As a result, most of the gas outside this 
region cools and converts to stars. This results in a small gaseous core. 
In contrast, in the galcon approach galaxies are treated more 
globally, with a wind and energy feedback prescription that spreads 
energy over an extended region. This feature is the main ingredient 
leading to the formation of an isothermal core in the GR cluster.
Energy feedback in our new approach is deposited directly into 
the gaseous layer surrounding the galaxy. The galcon feedback 
successfully spreads out heat energy into IC gas resulting in a large 
core in the GR cluster. As a result, the temperature profile 
of the GR overlaps with the observational data.

\label{fig:temperature}
\begin{figure}[h]
\centering
\epsfig{file=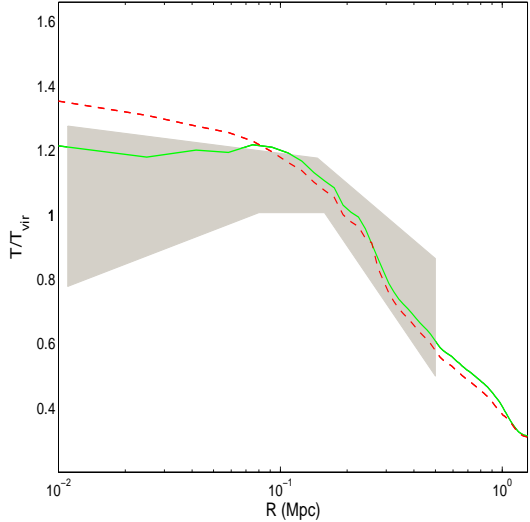,height=7cm,width=7cm,clip=}
\caption{Normalized temperature profiles of IC gas in the GR cluster, 
solid green line, and CR cluster, dashed red line. The shaded gray area
shows the mean temperature profile (including 1$\sigma$ uncertainties) 
from observations of 15 clusters (Pratt et al. 2007).}
\end{figure}

\subsection{Gas metallicity}

One of the unique advantages of the new approach is its capability to 
trace not only the total metallicity in the IC gas, but also the unique 
metallicity enrichment history and its spatial distribution as result 
of each physical process that is included in the code. Fig 7 shows the 
evolution of metallicity in the two 
clusters. Recent high-resolution Chandra observations of a large sample 
of clusters (Balestra et al. 2007, Maughan et al. 2008) have shown a 
significant increase in the metallicity towards lower redshifts, as 
seen in the GR cluster. These results imply that the average IC 
metallicity at the current epoch is at least twice higher than its 
initial value around the time of cluster formation. The slope of the 
metallicity evolution in the GR cluster is consistent with these 
results. This behavior also agrees with that predicted by Ettori (2005) 
based on models of SFR and subsequent evolution of SN feedback. In 
contrast, the metallicity evolution in the CR is much weaker. This 
relatively flat behavior is probably caused due to the over-estimated 
SFR of star particles in the CR cluster at higher redshifts (see figure 
3. As discussed above, the SFR in the CR cluster at its formation 
($z <2$) is substantially higher than expected, and a higher fraction 
of metals is ejected into IC space already at higher redshifts, 
flattening the metallicity evolution.

\label{fig:metal_z}
\begin{figure}[h]
\centering
\epsfig{file=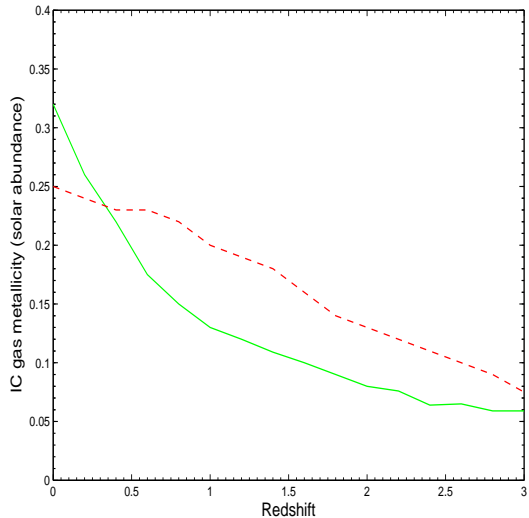,
height=7cm,width=7cm,clip=}
\caption{The mean IC metal abundance as a function of redshift for
the GR cluster, solid green line, and the CR cluster, dashed red line.}
\end{figure}

Metallicity in the GR cluster (Fig 
8) is due to 
enrichment by both winds and ram pressure stripping, with the former 
process being more effective at higher redshifts, since it is driven by 
shocks from SN that are then more prevalent. At early periods of 
cluster evolution a higher fraction of galaxies are outside the cluster 
core, where metals are preferentially deposited. As the cluster evolves, 
galaxies are more centrally distributed, so metals are more effectively 
spread in the central region. This results in an approximately constant 
metallicity across the cluster. On the other hand, because gas stripping 
depends on the local IC gas density, which builds up as the cluster 
evolves, the contribution to the metallicity is larger at lower $z$, 
and is more concentrated in the high density core, resulting in a 
substantial metallicity gradient.

The total metallicity in the GR cluster is roughly constant out to 
$\sim$700 kpc; it decreases at larger radii, with a mean value of 
0.32$Z_{\odot}$, which is in the observationally determined range, 
$(0.3-0.4)Z_{\odot}$ (e.g., Baldi et al. 2007). The mean metallicity 
in the CR cluster is 0.25$Z_{\odot}$, somewhat lower than typical. 
Moreover, its steep decline already in the central region is also 
inconsistent with observations, which show a nearly constant metallicity 
in the central few hundred kpc (Hayakawa et al. 2006, Pratt et al. 
2006), with the exception of a small galactic-size region at the 
cluster center where the metallicity is higher (Snowden et al. 2008). 
A much shallower gradient is observed in cooling flow clusters (e.g., 
De Grandi et al. 2004), but our simulated clusters have no cooling 
flows. We conclude that in our GR simulation - which includes galactic 
winds and ram pressure stripping - both the metallicity evolution, the 
level of metallicity and its spatial profile are consistent with 
observations, whereas neither property is well reproduced in the CR 
cluster.

\begin{figure}[h]
\label{fig:metallicity}
\centering
\epsfig{file=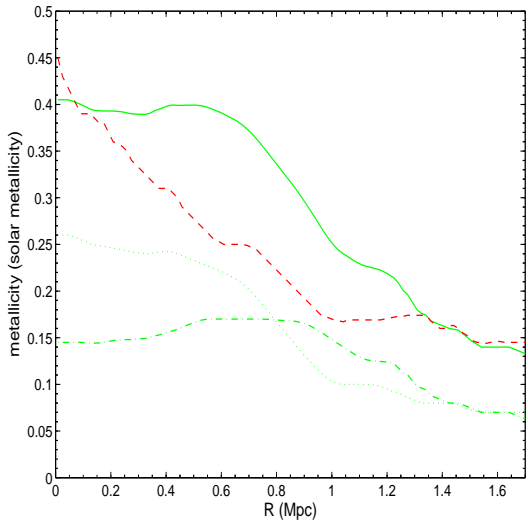,height=7cm,width=7cm,clip=}
\caption{Metallicity profile of the GR cluster at $z=0$, solid green 
line, compared to that in CR cluster, dotted red line. The total GR 
metallicity is the sum of the metallicity due to galactic winds (dashed-dotted 
green line) and the metallicity due to ram-pressure stripping (dashed green 
line).}
\end{figure}

In summary, feedback in our the new approach is shown to be very 
efficient in spreading energy deposition over the central region of the 
cluster, resulting in SF history and IC gas properties that are more 
consistent with observations. 

\subsection{Gas entropy profile}

The gas entropy is a basic quantity that could yield additional insight 
on the thermal evolution of the gas beyond what can be learned separately 
from the density and temperature. Defined as 
\begin{equation}
K=k_B T n^{-2/3} \; ,
\end{equation}
where $k_B$ is the Bolzmann constant and $n$ is the gas number density, 
the entropy is the simplest combination of $n$ and $T$ that is invariant 
under adiabatic processes in the gas. Both the level and radial profile 
of the entropy are useful for determining the effectiveness and 
distribution of non-gravitational heating of IC gas. Spherical accretion 
models of non-radiative clusters (Tozzi \& Norman 2001, Borgani et al. 
2002, Voit et al. 2005) predict a power law profile, $K(r) 
\propto r^{1.1}$ for $0.1-0.2\leq r/r_{200} \leq 1.0$. Within 
$0.1 r_{200}$

a gradual flattening of the profile (with a large scatter in the central 
value of $K$) has been determined (Pratt et al. 2006, Donahue 
et al. 2006, Vikhlinin et al. 2006). If further confirmed by more 
sensitive, high resolution measurements that do indeed allow detailed 
modeling of both the density and temperature profiles in the inner 
core regions, the apparent entropy flattening - 
referred to as entropy `floor' - could possibly be a result of heating 
processes that occurred in the early stages of the (hierarchical) 
cluster formation, or perhaps due to substantial enhancement of 
feedback processes late in the formation stage.

In many cluster simulations cluster cores are too compact, and with 
increasing temperature profile towards the center. This is likely 
due to inadequate implementation of the processes of galactic winds 
and ram pressure stripping, which are less important in the outer 
regions of clusters, so their impact in not seen in outer cluster 
regions, where simulations succeed in reproducing the entropy 
profile at $r > 0.1 r_{200}$ (e.g., Nagai et al. 2006). This possibly 
is the reason why most simulations fail to reproduce the central 
entropy profile - it flattens too quickly at a level that 
is higher than indicated by observations. The high entropy floor 
problem occurs in both AMR and SPH codes, but is known to be less 
severe in SPH simulations (e.g., Voit et al. 2005), in which the 
temperature profile increases only slightly towards the center.
We note that in a recent analysis of X-ray and lensing measurements of 
A1689 Lemze et al. (2008) have found the `entropy floor' to be even 
lower than the previously determined level, further increasing the 
discrepancy between simulations and observations.

\label{fig:entropy}
\begin{figure}[h]
\centering
\epsfig{file=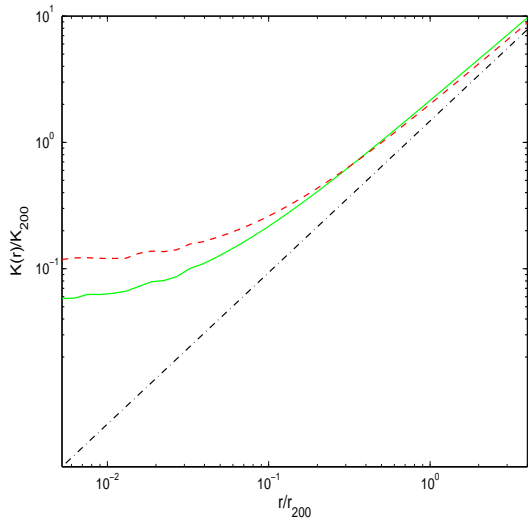,height=7cm,width=7cm,clip=}
\caption{Dimensionless entropy $K/K_{200}$ as function of scale radius
$r/r_{200}$ for the GR cluster, solid green line, and the CR cluster, 
dashed red line. The analytic formula of Voit et al. (2005) for the 
entropy profile in non-radiative AMR clusters is also shown by the 
dashed-dotted black line.}
\end{figure}

\noindent 
We compare the behavior of the entropy profiles in our two simulated 
clusters in Fig 9. In order to best clarify the 
behavior of the entropy, we subtracted the massive clump placed at 
$r \sim 600$ kpc from the center from both clusters. As can be expected 
from the previously shown density and temperature profiles, flattening 
of the entropy profile in CR cluster occurs at larger radius as 
compared to that of the GR cluster. The combination of a larger 
gas core radius and the isothermal temperature profile in the GR 
cluster yields a relatively low entropy floor. The entropy 
continues to decrease even below $0.1 r_{200}$; it gradually flattens 
only below $0.03 r_{200}$, in accord with the trend indicated by 
observations (e.g., Cavagnolo et al. 2009).

\section{Conclusion}
\label{conclusion}

In this paper we described our galcon approach to simulating baryonic 
processes in clusters using the new {\em Enzo-Galcon} AMR code. As we 
have demonstrated, the combination of galcons with improved 
semi-analytic modeling of the relevant baryonic processes yields a 
powerful code that is capable of reproducing the basic properties of 
clusters. This novel approach successfully describes SF and the basic 
properties of IC gas, by a more realistic implementation of galactic 
wind and ram-pressure stripping of metal-enriched IS media into IC space.

With improved measurement quality in all the major frequency bands 
(microwave, optical, and X-ray), there is a great need to expand 
simulation codes, both in the range of physical processes that are 
taken explicitly into account, and in the capability to attain higher 
spatial resolution. Towards this goal we will use a new and improved 
version (1.5) of the {\em Enzo} code. We expect that the combined 
{\em Enzo-Galcon} code will make it possible to resolve scales to a 
typical accretion radius, or $\sim 2-3$ kpc, an important milestone 
that is required for a more meaningful physical description of the 
interaction of a cluster galaxy with ambient DM and IC gas. With the 
upgraded code we will be able to identify a much larger number of 
halos, to replace galactic halos with galcons as the cluster evolves 
(rather than doing so only once at an early time), to include 
additional feedback sources (such as AGN), to identify galactic 
merger sites, and to improve our prescription for how mass and 
energy are deposited into IC space. We will also have better 
capability to simulate more realistically the properties of DM 
since the identification of a relatively large number of galaxies, 
and the ability to follow their feedback processes, translate also 
to a more physical description of the impact of galaxies and their 
feedback on the DM, especially in the central cluster region. By 
comparing similar properties of galaxies, IC gas, and DM, we are 
likely to gain additional insight on all these important cluster 
components.

\vspace*{1cm}
We thank Alexei Kritsuk and Brian O'Shea for many useful discussions, 
and Andrew Hopkins for providing the observational SFR data. 
Work at Tel Aviv University is supported by US-IL Binational Science 
foundation grant XXX/09. The simulations were performed on the Data 
Star system at the San Diego Supercomputer Center using LRAC allocation 
TG-MCA98N020. This work has been partially supported by NSF grants 
AST-0708960 and AST-0808184 to MLN.

\newpage
\def\ref{\par\noindent\hangindent 20pt}
\noindent
{\bf References}
\ref Allen S.W. \& Fabian A.C. 1998, \mn 297, 63.
\ref Arieli Y. \& Rephaeli Y. 2003, \na 8, 517A.
\ref Arnaud M. \& Evrard A.E. 1999, \mn 305, 631.
\ref Baldi A. et al. 2007, \apj 666, 835.
\ref Balestra I. et al. 2007, \aa 462, 429.
\ref Balogh M.L. et al. 2001, \mn 326, 1228.
\ref Berger M. J. \& Colella P. 1989, J. Comput. Phys., 82, 64.
\ref Boleman, J., Dogiel, V. A. \& Ptuskin, V. S., 1993 \aa 267, 372.
\ref Borgani S. et al. 2002, \ph0310794.
\ref Borgani S. et al. 2008, SSRv 134, 269.
\ref Borgani S. et al. 2008, SSRv 134, 379.
\ref Bruggen M. \& Ruszkowski M. 2005, \ph0512148.
\ref Bryan G. L. \& Norman M. L. 1997, in ASP Conf. Ser. 123, Computational Astrophysics, ed. D. A. Clarke \& M. Fall (San Francisco: ASP), 363.
\ref Bryan G. L. \& Norman M. L. 1999, in Workshop on Structured Adaptive Mesh Refinement Grid Methods, ed. N. Chrisochoides (IMA Volumes in Mathematics 117), 165.
\ref Bryan G. L. Abel T. \& Norman M. L. 2001, in Supercomputing 2001 (IEEE, http://www.sc2001.org/papers/).
\ref Castillo-Morales A. \& Schindler S. 2003, \aa 403, 433.
\ref Cavagnolo K.W et al. 2009, \apjs 182, 12.
\ref Cavaliere A., Menci N. \& Tozzi P. 1998, \apj 501, 493. 
\ref Cavaliere A., Lapi A. \& Menci N. 2002, \apj 581, L1.
\ref Cen R. \& Ostriker J.P. 1992, \apjl 399, L113.
\ref Chandrasekhar S. 1943, \apj 97, 255.
\ref Colin P., Klypin A. \& Kravtsov A.V. 2000, \apj 539, 561.
\ref Connolly A.J. et al. 1997, \apj 486, L11.
\ref de Blok W.J.G. et al. 2001, \aj 122, 2396.
\ref De Young D.S. 1978, \apj 223, 47.
\ref Domainko, W., et al. 2006, \aa, 452, 795.
\ref Efstathiou G., Davis M. White S. D. M., \& Frenk C. S. 1985, \apj, 57, 241.
\ref Eisenstein D. J. \& Hut P. 1998, \apj, 498, 137.
\ref Ettori S., De Grandi S. \& Molandi S. 2002, \aa 391, 841.
\ref Ettori S. 2005, \mn 362, 110.
\ref Fabian A. et al. 2000, \mn 318, L65. 
\ref Finoguenov A., Reiprich T. \& Bohringer H. 2001, \aa 368, 769.
\ref Frenk C.S. et al. 1999, \apj 525, 554.
\ref Fukazawa Y., Kawano N. \& Kawashima K. 2004, \apj 606, L109.
\ref Gavazzi R. et al. 2003, \aa 403, 11.
\ref Genadin N.Y. 1998, \mn 294, 407.
\ref Giavalisco M. et al. 2004, \apj 600, L103.
\ref Gunn J.E. \& Gott J.R. 1972, \apj 176, 1.
\ref Hansen S.A \& Moore B. 2006, \na 11, 333.
\ref Harnquist L. \& Springel V. 2003, \mn 341, 1253.
\ref Hayakawa A. et al. 2004, \apj 56, 743.
\ref Helsdon S.F. \& Ponman T.J. 2000, \mn 315, 356.
\ref Hopkins A.M. \& Beacom J.F. 2006, \apj 651, 142.
\ref Holden B. et al. 2002, \aj 124, 33.
\ref Host et al. 2009, \apj 690, 358.
\ref Jing Y.P. \& Suto Y. 2000, \apj 529, 69.
\ref Kay et al. 2007, \mn 377, 317.
\ref Kapferer W. et al. 2005, \ph0503559.
\ref Kapferer W. et al. 2007, \aa 466, 813.
\ref Katz N., Weinberg D.H. \& Hernquist L. 1996, \apjs 105, 19.
\ref Klypin A., Gottlober S. \& Kravtzov A.V. 1999, \apj 516, 530.
\ref Klypin A. et al. 2001, \apj 554, 903.
\ref Learche I. \& Schlickeiser R. 1980, \apj 239, 1089.
\ref Lemze D. et al. 2008, \mn 386, 1092.
\ref Lemze D. et al. 2009, \ph0810.3129.
\ref Lilly S.J. et al. 1996, \apj 460, L1.
\ref Madau P.  et al. 1996, \mn 283, 1388.
\ref Maughan B.J et al. 2008, \apjs 174, 117.
\ref Markevitch M. 1998, \apj 504, 27. 
\ref McNamara B.R. et al. 2000 \apj 534, L135.
\ref Menci N. et al. 2002, \apj 575, 18.
\ref Metzler C.A. \& Evrard A.E. 1994, \apj 437, 564.
\ref Moore B. et al. 1998, \apj 499, 5.
\ref Nagai D., Vikhlinin A. \& Kravtsov A.V. 2007, \apj 655, 98.
\ref Nagamine K. 2002, \apj 564, 73.
\ref Nagamine K. et al. 2004, \apj 610, 45.
\ref Navarro J. F., Frenk C. S. \& White S. D. M. 1997, \apj 490, 493.
\ref Nevalainen J., Markevitch M. \& Forman W. 2000, \apj 532, 694.
\ref Norman M. L. \& Bryan G. L. 1999, in Numerical Astrophysics: Proc. Int. Conf. on Numerical Astrophysics 1998 (NAP98), ed. S. M. Miyama, K. Tomisaka, \& T. Hanawa ( Dordrecht: Kluwer), 19.
\ref Novicki M.C., Sornig M. \& Henry J.P. 2002, \aj 124, 2413.
\ref Pettini et al. 2001, \apj 554, 981.
\ref Pettini et al. 2002, \apj 569, 742.
\ref Ponman T.J., Cannon D.B. \& Navarro J.F. 1998, \ph9810359.
\ref Ponman T.J., Sanderson J.R. \& Finogunenov A. 2003, \mn 343, 331.
\ref Pratt et al. 2007, \aa 461, 71.
\ref O'shea B. et al. 2004, \ph0403044.
\ref Ouchi M. et al. 2004, \ph0309655.
\ref Rasia E. G. \& Moscardini L. 2003, \ph0309405.
\ref Rephaeli Y. 1979, \apj 227, 364.
\ref Rephaeli Y. \& Salpeter 1980, \apj 240, 20.
\ref Rephaeli Y., Gruber D. \& Arieli Y. 2005, \apj, submitted. 
\ref Sand D.J. et al. 2002, \apj 574, 129. 
\ref Sanders J.S. et al. 2004, \mn 349, 952.
\ref Schindler S. et al. 2005, \ph0504068.
\ref Sijacki D. \& Springel V. 2006, \mn 366, 397.
\ref Somerville R.S., Primack J.R. \& Faber S.M. 2001, \mn 320, 504.
\ref Spergel D.N. et.al. 2006, \ph0603449.
\ref Steidel C.C. et al. 1996, \apj 462, L17.
\ref Strong A.W. \& Mattox, J. R. 1996 \aa 308, L21.
\ref Strong A.W. \& Moskalenko I.V. 1998, \ph9807150.
\ref Sun M. et al. 2009, \apj 693, 1142.
\ref Syer D. \& White S.D.M. 1998, \mn 293, 337.
\ref Tornatore L. et al. 2007, \mn 382, 1050.
\ref Torres D. 2004, \ph0407240.
\ref Tozzi P. \& Norman C. 2001, \apj 546, 63.
\ref Valageas P. \& Silk J. 1999, \aa 350, 725. 
\ref van den Bosh F.C. \& Swaters R.A. 2001, \mn 325, 1017.
\ref Voit G.M. 2005, \rmp 77, 207.
\ref Wu X.P. \& Xue Y.J. 2002, \apj 569, 112.
\ref Zirakashvili, V. N., Breitschwerdt, D., Ptuskin, V. S. \& Volk, H. J. 1996, \aa 311, 113.
\ref Zeldovich, Y. B. 1970, \aa, 5, 84.

\end{document}